\documentclass[usenatbib,onecolumn]{mn2e}
\usepackage{natbibmnfix,graphicx,times}
\def\bdm{\begin{displaymath}}
\def\edm{\end{displaymath}}

\newcommand{\Mpc}{\mbox{ Mpc}}

\newcommand{\MHz}{\mbox{ MHz}}

\newcommand{\kb}{k_{\rm B}}

\newcommand{\bxhi}{\bar{x}_{\rm HI}}
\newcommand{\xhi}{x_{\rm HI}}

\newcommand{\fcoll}{f_{\rm coll}}

\newcommand{\deriv}{{\rm d}}

\title[The 21 cm-CMB cross correlation]{Reionization and the large-scale 21 cm-cosmic microwave background cross correlation}

\author[Adshead \& Furlanetto]{Peter J. Adshead\thanks{peter.adshead@yale.edu} \& Steven R.  Furlanetto \\
Yale Center for Astronomy and Astrophysics, Yale University, PO Box 208121, New Haven, CT 06520-8121}

\begin{document}

\maketitle

\begin{abstract}
Of the many probes of reionization, the 21 cm line and the cosmic
microwave background (CMB) are among the most effective.  We examine
how the cross-correlation of the 21 cm brightness and the CMB
Doppler fluctuations on large angular scales can be used to study
this epoch. We employ a new model of the growth of large scale
fluctuations of the ionized fraction as reionization proceeds. We
take into account the peculiar velocity field of baryons and show
that its effect on the cross correlation can be interpreted as a
mixing of Fourier modes. We find that the cross-correlation signal
is strongly peaked toward the end of reionization and that the sign
of the correlation should be positive because of the inhomogeneity
inherent to reionization. The signal peaks at degree scales
($\ell\sim100$) and comes almost entirely from large physical scales
($k\sim10^{-2}$ Mpc). Since many of the foregrounds and noise that
plague low frequency radio observations will not correlate with CMB
measurements, the cross correlation might appear to provide a robust
diagnostic of the cosmological origin of the 21 cm radiation around
the epoch of reionization. Unfortunately, we show that these signals
are actually only weakly correlated and that cosmic variance
dominates the error budget of any attempted detection. We conclude
that the detection of a cross-correlation peak at degree-size
angular scales is unlikely even with ideal experiments.
\end{abstract}


\section{Introduction}

During some epoch between recombination ($z\sim1100$) and today, the
intergalactic medium (IGM) underwent a transformation from almost
completely neutral to almost completely ionized. When and how the
IGM was ionized remains one of the most exciting open questions in
cosmology \citep{Barkana:2000fd, Furlanetto:2006jb}. The timing and
duration of this epoch of reionization contains a wealth of
information about the first cosmic structures, information which is
expected to help explain how the primordial density perturbations
observed in the cosmic microwave background (CMB) evolved into the
complex structures we observe in the low redshift universe today.

Currently there are only weak observational constraints on the epoch
of reionization. The absence of a Gunn-Peterson trough in the
spectra of $z<6$ quasars indicates that reionization was complete by
$z\sim6$ \citep{fan02,white03,fan06}, while integral constraints on
the total optical depth from WMAP imply reionization began at
$z\geq10$ \citep{page06,Spergel:2006hy}. A variety of other
methods have been proposed, but none impose strong constraints
(for recent summaries, see \citealt{Fan:2006dp} and
\citealt{Furlanetto:2006jb}).

The 21 cm hyperfine spin-flip transition of HI is the most exciting
prospective tracer of the cosmic gas before and during reionization.
This signal does not require the existence of bright background
sources, which may be rare at high redshifts, making the entire
epoch of reionization available
\citep{scott90,madau97,Furlanetto:2006jb}. It is a line transition,
so that observations at a given frequency select out a unique slice
of the high $z$ universe. Furthermore, fluctuations in the
brightness of the 21 cm signal are caused by density (and
ionization) inhomogeneities on all scales, making it a direct tracer
of the underlying matter distribution \citep{Zaldarriaga:2003du}.
Thus the spectral and angular variations of the 21 cm brightness
allow us to reconstruct a 3D map of reionization. The epoch of
reionization spans the formation of the first luminous sources,
which then ionized the surrounding gas \citep{Barkana:2000fd}. Thus
a 3D map of the evolution of the neutral fraction and density field
would provide an unprecedented view of this epoch of structure
formation.

Ionization of the neutral IGM creates free electrons off of which
CMB photons may be scattered. This Thomson scattering process has
several effects \citep{hu94,dodelson95}. Photons from multiple lines of sight are blended
together, damping the primary CMB anisotropies. The scattered
photons also gain some of the peculiar momentum of the free
electrons, generating a secondary anisotropy (most recently examined by \citealt{Giannantonio:2007za}). Finally, the
polarization dependence of Thomson scattering generates a new large
scale polarization from the anisotropic CMB photon field \citep{zalda97-reion}.

Taken by itself, the CMB temperature provides only an integral
constraint on the column density of ionized electrons, because the
net damping is nearly independent of the location of the electrons.
The induced polarization contains somewhat more information about the
history of reionization, but it is difficult to extract even in a
cosmic variance-limited survey \citep{holder03, mortonson07}.  Cross
correlating with the 21 cm signal should allow more information to be
extracted from both measurements. The cross correlation signal
arises from the velocity field of ionized baryons, which sources the
Doppler anisotropies in the CMB and also traces the linear
overdensity of neutral hydrogen, the source of the 21 cm brightness.
\citet{Alvarez:2005sa} showed that combining the two measurements
could in principle allow the extraction of the global reionization
history, which is otherwise difficult to measure (e.g.,
\citealt{shaver99,Furlanetto:2006tf}). One reason is that the signal from the 21 cm
radiation will be dominated by foregrounds and
detector noise which makes extracting useful information difficult.
Although many of these same foregrounds appear in the CMB, they are
much smaller and more easily removed (and of course the thermal
noise is uncorrelated).  Cross-correlation with the CMB could
therefore dramatically improve our confidence in the cosmological
origin of the 21 cm signal.

This work is an improved calculation of results presented by
\citet{Alvarez:2005sa}. In particular, we include the corrections to
the 21 cm brightness due to redshift space distortions caused by the
peculiar velocity field of neutral hydrogen (as in
\citealt{Bharadwaj:2004nr}). We employ a more physically motivated
reionization history together with a new model of the growth of the
ionized contrast as reionization proceeds. We also re-examine the
prospects for detecting the large scale cross-correlation.
Unfortunately, we find that it will be extremely difficult because
of cosmic variance. Throughout this work, we will confine the
discussion to large angular scales ($\sim$degree) where the details
of the the ionized bubbles that appear during reionization
\citep{Furlanetto:2004nh} largely average out.  On smaller scales,
these arcminute structures will also induce interesting
cross-correlations (e.g.,
\citealt{Cooray:2004ei,Salvaterra:2005js,Slosar:2007sy}), but the
simple models based on linear theory used here will not suffice to
describe them.  Also, note that we will assume gaussian fluctuations
throughout, as appropriate for linear matter fluctuations on the
large scales we study.  Only on smaller scales will higher-order
correlations become important \citep{Cooray:2004ei}.

This paper is organized as follows. In \S 2 expressions for the 21
cm brightness, CMB Doppler brightness, and the cross-correlation
signals are derived. Equation (\ref{21-Dopp}) is the main result.
Section 3 details the reionization model, and \S 4 presents
calculations of the observability of our predictions for future
radio telescopes. Finally, we conclude in \S 5.

Throughout our work the following Fourier convention is used
\begin{equation}
f(\hat{{\bf n}}, \eta) = \int\frac{\deriv^{3}{\bf
k}}{(2\pi)^{3}}f_{{\bf k}} e^{-i{\bf k}\cdot\hat{{\bf
n}}(\eta_{0}-\eta)}.
\end{equation}

The cosmological parameters are set at $\Omega_{0}=0.24$,
$\Omega_{b} = 0.042$, $\Omega_{\Lambda} = 0.76$, $h=0.73$,
$\sigma_{8} = 0.8$ and $n=0.96$, consistent with the WMAP year three
data \citep{Spergel:2006hy}, and we use the matter power spectrum of
\citet{Eisenstein:1997jh}. We work in natural units where $c=1$, so
that in a flat geometry the conformal time,
\begin{equation}
\eta_{0}-\eta(z) = \int_{0}^{z}\frac{\deriv z'}{H(z')},
\end{equation}
equals the comoving distance, $r(z) = \eta_{0}-\eta(z)$. Throughout
$H(z)$ denotes the Hubble parameter. We will also use the expansion
of the plane wave,
\begin{eqnarray}
e^{-i {\bf k}\cdot {\bf x}} & = &
\sum_{\ell=0}^{\infty}(-i)^{\ell}(2\ell
+1)j_{\ell}(kx)\mathcal{P}_{\ell}(\hat{{\bf k}}\cdot \hat{{\bf x}})\\
& = & 4\pi \sum_{\ell, m}(-i)^{\ell}j_{\ell}(kx)Y_{\ell m}(\hat{{\bf
k}})Y_{\ell m}^{*}(\hat{{\bf x}}) \label{plwvexp},
\end{eqnarray}
where ${\mathcal P}_\ell$ is the Legendre polynomial of order $\ell$.

\section{21 cm and CMB doppler Temperature Anisotropy}

\subsection{21 cm Temperature Anisotropy}

As long as the spin or excitation temperature, $T_{s}$, of the 21 cm
transition in a region of the intergalactic medium (IGM) differs
from the CMB temperature that region will appear in either emission
($T_{s} > T_{cmb}$) or absorption ($T_{s}< T_{cmb}$) when viewed
against the CMB. Variations in the density of neutral hydrogen would
appear as fluctuations in the sky brightness of this transition
\citep{scott90,madau97,Furlanetto:2006jb}.

The optical depth of the IGM in the hyperfine transition is
\citep{field58}
\begin{equation}
\tau_{21} = \frac{3 c^{3}\hbar A_{10} n_{H}}{16
\kb\nu_{0}^{2}T_{s}(1+z)(dv_{r}/dr)}\xhi,
\end{equation}
where $A_{10}=2.85 \times 10^{-15}$ s$^{-1}$ is the spontaneous
emission coeffiecient, $\kb$ is the Boltzmann constant, $n_H$ is
the number density of hydrogen, and $\xhi$ is the neutral fraction.
$dv_{r}/dr$ is the gradient of the radial
velocity along the line of sight, $v_{r}$ being the physical
velocity and $r$ the comoving distance.  This factor accounts for the bulk motion of the hydrogen,
which causes a Doppler shift from the intrinsic line frequency,
$\nu_{0} = 1420.2$ MHz.

Peculiar velocities induce a small perturbation in the gradient of
the proper velocity away from the Hubble flow
\begin{equation}
\frac{\deriv v_{r}}{\deriv r} = a(z)H(z) + \frac{\partial
v}{\partial r},
\end{equation}
where $v$ is the radial peculiar velocity and $a(z)$ is the scale factor.
To first order in the density perturbations the optical depth in the
hyperfine transition is then
\begin{equation}\label{tau21}
\tau_{\rm 21} \approx 8.6 \times
10^{-3}\xhi\left(1+\delta_{b}(\hat{{\bf n}}, \eta) -\frac{1}{a(z)
H(z)}\frac{\partial v_{r}}{\partial
r}\right)\left[\frac{T_{cmb}(z)}{T_{s}}\right]\left(
\frac{\Omega_{b}h^{2}}{0.02}\right)\left[\left(\frac{0.15}{\Omega_{m}h^{2}}
\right)\left(\frac{1+z}{10}\right)\right]^{1/2},
\end{equation}
where $T_{cmb}(z)=2.725(1+z)$ K and we have assumed that the
universe is matter-dominated at the redshifts of interest.
In arriving at this expression we have neglected a number of terms,
such as the effect of the CMB dipole in the gas frame and the local
effects of gravitational redshifting, which also contribute to the
21 cm optical depth. However, these additional
terms contribute less than $1\%$ to the anisotropy on the scales of
interest, and equation (\ref{tau21}) suffices for our purposes \citep{Lewis:2007kz}.

Following \citet{Zaldarriaga:2003du}, we define $T_{21}(\hat{{\bf
n}}, z)$ to be the observed brightness temperature increment between
this patch and the CMB at an observed frequency $\nu$ corresponding
to a redshift $1+z = \nu_{0}/\nu$ in a direction $\hat{{\bf n}}$,
\begin{eqnarray}
T_{ 21}( z) & \approx & \frac{1}{1+z}(T_{s}-T_{cmb})\tau_{ 21},
\end{eqnarray}
where we have assumed that $\tau_{21}$ is small.

We now write the observed differential brightness temperature of the
21 cm emission line at $\lambda = 21$ cm$(1+z)$ in the direction
$\hat{{\bf n}}$ as
\begin{equation}
T_{21}(\hat{{\bf n}}, z) =
T_{0}(z)\int_{0}^{\eta_{0}}\deriv\eta'W[\eta(z)-\eta']\psi_{21}(\hat{{\bf
n}}, \eta'),
\end{equation}
where $W[\eta(z)-\eta']$ is a normalized
($\int_{-\infty}^{\infty}\deriv x W[x]=1$) spectral response of an
instrument which is centered at $\eta(z)-\eta' = 0$.  $T_{0}(z)$ is
a normalization factor given by
\begin{equation}
T_{0}(z)\simeq
23~\textrm{mK}~\left(\frac{\Omega_{b}h^{2}}{0.02}\right)\left[\left(\frac{0.15}{\Omega_{m}h^{2}}\right)
\left(\frac{1+z}{10}\right)\right]^{1/2},
\end{equation}
and
\begin{equation} \label{psi}
\psi_{21}(\hat{{\bf n}}, \eta)\equiv \xhi(\hat{{\bf n}},
\eta)\left[1+\delta_{b}(\hat{{\bf n}}, \eta)-\frac{1}{a(\eta)
H(\eta)}\frac{\partial v_{r}}{\partial
r}\right]\left[1-\frac{T_{cmb}(\eta)}{T_{s}(\hat{{\bf n}},
\eta)}\right].
\end{equation}
We assume that the spectral resolution of the instrument is much
smaller than the features of the target signal in redshift space. We
thus set $W[x]=\delta^{D}(x)$, to obtain
\begin{equation}\label{window}
T_{21}(\hat{{\bf n}},z)=T_{0}(z)\psi_{21}[\hat{{\bf n}},
\eta(z)].
\end{equation}
On very small scales, this is a poor approximation because the finite
bandwidth averages over many oscillations, damping the signal.
However, for our regime of interest at $\ell\sim100$ this is
expected to be an excellent approximation
\citep{Zaldarriaga:2003du}.

During most of reionization, we expect $T_{s}\gg T_{cmb}$
\citep{ciardi03-21cm,Furlanetto:2006tf}, so we neglect the $T_{cmb}/T_{s}$ term in
equation (\ref{psi}). By writing the ionized fraction contrast
\begin{equation}
\delta_{i}\equiv \frac{x_{i}-\bar{x}_{i}}{\bar{x}_{i}},
\end{equation}
where $x_{i}$ is the ionized fraction and the overbar denotes an average quantity,
to first order in the density perturbations, equation (\ref{psi}) can
be written
\begin{equation}\label{psi22}
\psi_{21}(\hat{n}, \eta) = \bxhi\left(1+\delta_{b}(\hat{n},
\eta)-\frac{1}{a(z) H(z)}\frac{\partial v_{r}}{\partial
r}\right)-\bar{x}_{i}\delta_{i}(\hat{n}, \eta).
\end{equation}
It is important to note here that this is not necessarily a
well-defined perturbation expansion. At any given point in space,
the ionized fraction is either zero or unity, so that averaged over
sufficiently small scales (of order the characteristic size of the
HII regions), $\delta_{i}$ itself will be at least of order unity.  However,
we will only examine large scales ($\gg 10 \Mpc)$ which average over
many of these highly ionized regions, so equation~(\ref{psi22}) is
acceptable.

Working in Fourier space, the baryon velocity field is related to
the density contrast via the continuity equation, ${\bf v}_{{\bf
k}}=-i{\bf k}/k^{2}\dot\delta_{b,{\bf k}}(\eta)$, where the overdot
denotes a derivative with respect to conformal time. At late times
on the scales of interest, the perturbations scale as the linear
growth factor $D_{1}$, so $ \dot{\delta}_{b,{\bf
k}}(\eta)=\delta_{b,{\bf k}}a(\eta)H(\eta)f$, where $f\equiv
\deriv\ln D_{1}/\deriv \ln a$ is the dimensionless linear growth
rate or redshift space distortion factor \citep{Kaiser:1987qv}.
Equation (\ref{psi22}) becomes
\begin{equation}\label{alt2}
    \psi(\hat{{\bf n}},
    \eta(z))=\int\frac{\deriv^{3}k}{(2\pi)^{3}}\left[\bxhi(z)\delta_{{\bf k}}(z)
    \left(1-\frac{f}{k^{2}}\frac{\partial^{2}}{\partial
    \eta^{2}}\right)
    -\bar{x}_{i}(z)\delta_{i{\bf k}}(z)\right]e^{i k
    \mu\eta}.
\end{equation}
The multipoles of the spherical harmonic expansion are then
\begin{equation} \label{21-multipoles}
a_{\ell m}^{21}(z) =
4\pi(-i)^{\ell}T_{0}(z)\int\frac{d^{3}k}{(2\pi)^{3}}\left\{\bxhi(z)\delta_{{\bf
k}}(z)\right.\left.J_{\ell}[k(\eta_{0}-\eta(z))]
-\bar{x}_{i}(z)\delta_{i {\bf
k}}(z)j_{\ell}[k(\eta_{0}-\eta(z))]\right\} Y_{\ell m}^{*}(\hat{{\bf
k}}).\nonumber
\end{equation}
where \citep{Bharadwaj:2004nr, Barkana:2004zy},
\begin{equation}\label{bharad}
J_{\ell}(x)\nonumber
=-f\frac{\ell(\ell-1)}{4\ell^{2}-1}j_{\ell-2}\left[k(\eta_{0}-\eta(z))\right]
    +
   \left(f\frac{(2\ell^{2}+2\ell-1)}{4\ell^{2}+4\ell-3}+1\right)j_{\ell}\left[k(\eta_{0}-\eta(z))\right]\\
  -f\frac{(\ell+2)(\ell+1)}{(2\ell+1)(2\ell+3)}j_{\ell+2}\left[k(\eta_{0}-\eta(z))\right],
\end{equation}
although we also include the redshift space distortion factor $f$.
This is well approximated by $f = \Omega_{m}^{0.6}$, which for the
redshifts of interest is approximately one. Neglecting the peculiar
velocity perturbations amounts to taking $J_{\ell}(x) \rightarrow
j_{\ell}(x)$.

To this point, our calculation is similar to that of
\citet{Alvarez:2005sa}, except that we have incorporated the
velocity correction in a different manner. They differentiated the
plane wave before expanding in spherical coordinates. This
left a factor $(1+\mu^2)$, where
$\mu = \hat{\bf n}\cdot{\bf k}/|{\bf k}|$ is the angle between the line
of sight direction $\hat{{\bf n}}$ and the photon propagation vector
${\bf k}$. On the other hand, we expand the plane wave into a
spherical basis before taking the derivative. Both approaches are
mathematically equivalent, but the advantage of ours will become
apparent when calculating the cross-correlation.

\subsection{The CMB Doppler Signal}

In Fourier space, the secondary
CMB temperature anisotropy from the Doppler effect is given by the
line of sight integral
\begin{equation}\label{Doppler}
T_{D}(\hat{n}, {\bf k}) = -T_{cmb}\int_{0}^{\eta_{0}}\deriv\eta
(-\dot{\tau})\mu\,v_{b {\bf k}}~ e^{i k\mu(\eta-\eta_{0})}e^{-\tau}=
-T_{cmb}\int_{0}^{\eta_{0}}\deriv\eta(-\dot\tau) e^{-\tau}\frac{v_{b
{\bf k}}}{i k}\frac{\partial}{\partial \eta}\left[e^{i
k\mu(\eta-\eta_{0})}\right].
\end{equation}
We define $\tau(\eta) =
\int_{\eta}^{\eta_{0}}\deriv\eta~\sigma_{T}n_{e}a$ so that
$\dot{\tau} = -\sigma_{T}n_{e}a$. Note that with this definition
$(-\dot\tau)>0$.

As above, the baryon velocity field is related to the density
contrast via the continuity equation. Thus in real space, assuming
an observer positioned at the present day ($\eta = \eta_{0}$), we have
\begin{equation}\label{Doppler1}
T_{D}(\hat{n})=T_{cmb}\int_{0}^{\eta_{0}}\deriv\eta\dot D(-\dot\tau)
e^{-\tau}\int\frac{\deriv^{3}k}{(2\pi)^{3}}\frac{\delta_{k}}{k^{2}}\frac{\partial}{\partial\eta}e^{-i
k\mu(\eta_{0}-\eta)}.
\end{equation}
Expanding in spherical coordinates, the multipole moments are
\begin{equation}\label{Doppler2}
a_{\ell,m}^{D}=4\pi
T_{cmb}(-i)^{\ell}\int_{0}^{\eta_{0}}\deriv\eta\dot{D}(-\dot\tau)
e^{-\tau}\int\frac{\deriv^{3}k}{(2\pi)^{3}}\frac{\delta_{k}}{k^{2}}\left[\frac{\partial}{\partial
\eta}j_{\ell}[k(\eta_{0}-\eta)]\right]Y_{\ell,m}^{*}(\hat{{\bf k}}).
\end{equation}

\subsection{Doppler-21 cm Cross Correlation}

We now calculate the cross correlation power spectrum,
$C_{\ell}^{21-D} = \langle a_{\ell m}^{21}a_{\ell m}^{D*}\rangle$.
We define the 3D power spectrum, $ \langle\delta_{{\bf k}}(z)
\delta_{{\bf k}'}(z)\rangle =(2\pi)^{3} \delta^{D}({\bf k}-{\bf
k}')P_{\delta\delta}(k, z)$  and the cross-correlation power
spectrum between the ionized fraction and density,
$\langle\delta_{{\bf k}}(z)\delta_{i, {\bf k}'}(z)\rangle =
(2\pi)^{3}\delta^{D}({\bf k}-{\bf k}')P_{\delta i}(k, z)$. We have
\begin{eqnarray}
\langle a_{\ell m}^{21}(z)a_{\ell m}^{D*}\rangle & = & \nonumber
   - T_{0}(z)T_{cmb}\frac{2 D_{1}(z)}{\pi}\int \deriv k\left[\bar{x}_{HI}(z)
    P_{\delta\delta}(k)J_{\ell}[k(\eta_{0}-\eta(z))]\right.\\
    &{}&-\left.
    \bar{x}_{i}(z)P_{\delta i}(k)j_{\ell}[k(\eta_{0}-\eta(z))\right]
    \int_{0}^{\eta_{0}}\deriv\eta'\dot
    D_{1}(-\dot\tau) e^{-\tau}
    \frac{\partial}{\partial\eta'}j_{\ell}[k(\eta_{0}-\eta')].
\end{eqnarray}
We perform an integration by parts, and neglect the surface term
since at early times  $\tau \approx \infty$ and at late times
$\dot\tau(0)$, the scattering rate, is essentially zero. We obtain
\begin{eqnarray}\label{21-Dopp}\nonumber
\langle a_{\ell m}^{21}(z)a_{\ell m}^{D*}\rangle & = &
    -T_{0}(z)T_{cmb}\frac{2 D_{1}(z)}{\pi}\int_{0}^{\eta_{0}}\deriv\eta'
    \frac{\partial}{\partial\eta'}\left[\dot
    D_{1}(-\dot\tau) e^{-\tau}\right]\int \deriv k j_{\ell}[k(\eta_{0}-\eta')]\\
   &{}&\times\left\{ \bxhi(z)
    P_{\delta\delta}(k)J_{\ell}[k(\eta_{0}-\eta(z))]-\bar{x}_{e}(z)
    P_{i\delta}(k)j_{\ell}[k(\eta_{0}-\eta(z))]\right\}.
\end{eqnarray}
Compared to the expression of \citet{Alvarez:2005sa}, we have the
factor $J_{\ell}[k(\eta_{0}-\eta(z))]$ instead of
$4/3j_{\ell}[k(\eta_{0}-\eta(z))]$.  To obtain
equation~(\ref{21-Dopp}), we have simply required the orthogonality
of the spherical harmonics to evaluate $\int \deriv\Omega
Y_{\ell}(\mu, \phi)Y^{*}_{\ell'}(\mu,\phi)$. However, if the factor
$\mu^2$ is included instead, as in \citet{Alvarez:2005sa}, then one
must evaluate $\int \deriv\Omega \mathcal{P}_{2}(\mu) Y_{\ell}(\mu,
\phi)Y^{*}_{\ell'}(\mu,\phi)$, where $\mathcal{P}_{2}(\mu)$ is the
second Legendre polynomial. \citet{Alvarez:2005sa} approximated this
integral with its value for $\ell=0$, introducing the factor $4/3$.

Ignoring the velocity corrections to the 21 cm signal for a moment
[this amounts to taking $J_{\ell}(x) \rightarrow j_{\ell}(x)$], and
using the Limber approximation (proved in appendix A):
\begin{equation}\label{limber}
 \frac{2}{\pi}\int_{0}^{\infty}k^2{}\deriv k \frac{P(k)}{k^{2}}j_{\ell}(kr)j_{\ell}(kr')\approx
    P\left(k=\frac{\ell}{r}\right)\frac{\delta(r-r')}{\ell^{2}},
\end{equation}
valid for $\ell \gg 1$, yields
\begin{eqnarray}\label{limberapp}
\frac{\ell^{2}C_{\ell}^{21-D}}{2\pi} =  T_{0}(z)T_{cmb}
\frac{D_{1}(z)}{2\pi} \frac{\partial}{\partial\eta}\left(\dot
    D_{1}\dot\tau e^{-\tau}\right)\left[\bar{x}_{HI}(z)
    P_{\delta\delta}\left(\frac{\ell}{r(z)}\right)-\bar{x}_{i}
    (z)P_{i\delta}\left(\frac{\ell}{r(z)}\right)\right].
\end{eqnarray}
This form is identical to equation
(17) of \citet{Alvarez:2005sa}, except without their factor of
$4/3$. It follows from equation (\ref{limberapp}) that the cross
correlation power spectrum should roughly trace the shape of the
underlying matter power spectrum (at least for uniform reionization, or for $\delta_i \propto \delta$
as below). The matter power spectrum has a broad peak on the scale
of the horizon size at matter-radiation equality, $k_{eq}\simeq
0.009\Mpc^{-1}(\Omega_{m}h^{2}/0.128)$. Using the fact the the
conformal time is on the order of $10^{4}\Mpc$ for large $z$,
equation (\ref{limberapp}) implies that the cross correlation power
spectrum will have a peak at degrees scales or $\ell\sim kr \sim 100$. For
the rest of the paper we take $\ell_{peak} = 100$.

Equation (\ref{limberapp}) implies another important fact: there is
a one-to-one correspondence between $k$ and $\ell$, so each
multipole samples only one scale. In the exact case this is only
approximately true; the right hand side of equation (\ref{limber})
is not a true delta function but has some finite width, which means
that modes are mixed. The peculiar velocity corrections further mix
the modes as seen through the appearance of $j_{\ell+2}(kx)$ and
$j_{\ell-2}(kx)$ in equation (\ref{bharad}). These redshift space
distortions boost the signal by about 20\% at $\ell_{peak}$.

To obtain a numerical result we need a sensible model for the
reionization history. On the scales of interest, density
perturbations grow linearly. We therefore use a growth factor
normalized so that $D_{1}(z_{ N})=1$, or
\begin{eqnarray}
    \dot{D}_{1}&=&
    -H(z)\frac{\deriv}{\deriv z}\left(\frac{1+z_{N}}{1+z}\right).
\end{eqnarray}
Assuming that only hydrogen is ionized,
\begin{eqnarray}\nonumber
    -\dot{\tau} &=&
    \frac{\sigma_{T}\rho_{b0}}{m_{p}}(1-Y_{p})(1+z)^{2}\bar{x}_{i}(z)\\
    &=&0.0525H_{0}\Omega_{b}h(1+z)^{2}\bar{x}_{i}(z),
\end{eqnarray}
where $Y_{p}=0.24$ is the helium mass abundance. Taking
$e^{-\tau}\approx 1$, since $\tau \sim 0.1 $ \citep{Spergel:2006hy},
then gives \citep{Alvarez:2005sa}:
\begin{equation}\label{avalrez}
\frac{\partial}{\partial\eta}\left[\dot{D}(-\dot{\tau})e^{-\tau}\right]=-0.0525H_{0}\Omega_{b}h
(1+z_{N})H(z)\frac{\deriv}{\deriv z}\left[\bar{x}_{i}(z)H(z)\right].
\end{equation}
The combination $(1+z_{N})^{2}P(k,z_{N})$ leaves equations
(\ref{21-Dopp}) and (\ref{limberapp}) independent of the epoch of
normalization $z_{N}$ \citep{Hu:1995fq}.

As noted above, the shape of the cross correlation power spectrum
for fixed $z$ is determined entirely by the underlying matter power
spectrum. For fixed $\ell$ its variation with redshift is determined
primarily by the quantity $\partial/\partial\eta(\dot D\dot\tau
e^{-\tau})$, which from equation (\ref{avalrez}) depends on
$\deriv/\deriv z\left[\bar{x}_{i}(z) H(z)\right]$. This implies
another important fact: the cross-correlation vanishes when
$\deriv/\deriv z\left[\bar{x}_{i}(z) H(z)\right]$ is constant and
will be the largest where $\left[\bar{x}_{i}(z) H(z)\right]$ has the
greatest rate of change.  Models with faster rates of reionization and reionization occuring
at higher redshifts will lead to larger cross-correlations (see below).
This behavior comes directly from the CMB Doppler signal; unless the
ionized fraction is changing, the line of sight integral in equation
(\ref{Doppler1}) suffers severe cancelation due to the oscillatory
nature of the Bessel functions. Furthermore, this means that if
dilution from the Hubble expansion is faster than the reionization
rate, then the CMB will be on average blueshifted by reionization
(and vice versa). During reionization $\bar{x}_{i}$ will almost
definitely be increasing very rapidly, so redshifting will be the
dominant effect.

\begin{figure}
\begin{centering}
\includegraphics[width = 8cm]{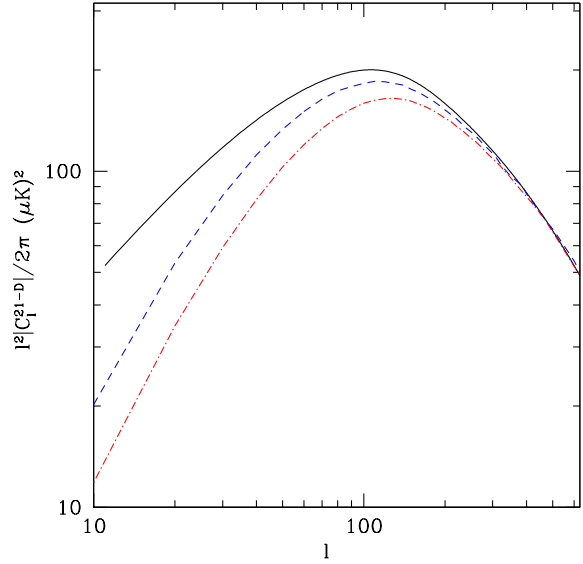}\\
\caption{Cross-correlation angular power spectrum. The  solid curve
is calculated using the Limber approximation. The dot-dashed curve
is the exact numerical integration without the velocity corrections
and the dashed curve includes the velocity corrections. We take
$z_{r}=15$, $\Delta z = 1$, and $z = 15$}\label{Transfer}
\end{centering}
\end{figure}

To get an idea of how the velocity corrections affect the signal we
chose a simple parametrization of the neutral fraction to compare
with previous results \citep{Furlanetto:2004ha, Alvarez:2005sa};
\begin{equation}\label{model}
\bxhi(z)=\frac{1}{1+\exp[-(z-z_{r})/\Delta z]},
\end{equation}
Figure~\ref{Transfer} shows the cross-correlation at $z=15$ for
homogeneous reionization ($P_{\delta i}=0$) using equation
(\ref{model}) with $\Delta z = 1$ and $z_{r}=15$. The three curves
show the Limber approximation of equation~(\ref{limber}), the exact
result, and the anisotropy ignoring velocity corrections (i.e., $J_l
\rightarrow j_l$), from top to bottom.  Note how the Limber
approximation works extremely well at $\ell \gg 100$ but
overestimates the signal by $\sim 10\%$ near $\ell_{peak}$ and fares
considerably worse at small $\ell$. Velocities boost the signal on
large angular scale by tens of percent but become unimportant on
small scales. Note that the result of \citet{Alvarez:2005sa}
corresponds to 4/3 times the lower curve, which at $\ell\sim 100$ is
not a bad approximation but gets worse as $\ell$ increases.

%Figure 1: Cross-Correlation Power

\subsection{The Cross-Correlation Coefficient}

We will find later on that it is convenient to re-cast many of our
results in terms of the cross-correlation coefficient $r$,
\begin{equation}
C_{\ell}^{21-D} = r\sqrt{C_{\ell}^{21}C_{\ell}^{CMB}},
\end{equation}
where $C_\ell^{21}$ is the angular power spectrum of the 21 cm signal and $C_\ell^{CMB}$ is the angular power spectrum of the CMB; in particular, we are interested in the Doppler contribution to the latter, $C_\ell^{DD}$.  The \emph{maximal} cross-correlation coefficient will be $r_{\rm max}=\sqrt{C_\ell^{DD}/C_\ell^{CMB}}$, because the 21 cm signal is (nearly) uncorrelated with the primordial CMB anisotropy and other secondaries.

We can estimate the intrinsic 21 cm signal with a little bit of work from equation (\ref{21-multipoles}). Neglecting redshift space distortions, which give about a $30\%$ correction at $\ell = 100$, and using the results above, we have \citep{Zaldarriaga:2003du}
\begin{eqnarray}
C_{\ell}^{21}(z)= (4\pi)^2 T_{0}(z)^{2} \int k^2 \frac{\deriv
k}{(2\pi)^{3}}j_{\ell}\left[k(\eta_{0}-\eta(z))\right]^{2}[\bxhi^2
P(k, z) + \bar{x}_i^2 P_{ii}(k,z) - 2 \bxhi \bar{x}_i P_{\delta
i}(k,z)], \label{cl21}
\end{eqnarray}
where $P_{ii}$ is the power spectrum of $\delta_i$.

We can estimate the CMB power from the Doppler signal from equation (\ref{Doppler2}),
\begin{equation}
\langle a_{\ell,m}^{D}a_{\ell,m}^{D~*}\rangle =
T_{cmb}^{2}\int_{0}^{\eta_{0}}\deriv\eta\frac{\partial}{\partial
\eta}\left[\dot{D}(-\dot\tau)
e^{-\tau}\right]\int_{0}^{\eta_{0}}\deriv\eta'\frac{\partial}{\partial
\eta'}\left[\dot{D}(-\dot\tau)
e^{-\tau}\right]
\frac{2}{\pi}\int \deriv k P(k)j_{\ell}[k(\eta_{0}-\eta)]j_{\ell}[k(\eta_{0}-\eta')],
\end{equation}
where we have used the orthogonality of the spherical harmonics and integrated
by parts twice. Using the Limber approximation (eq.~\ref{limberapp}), we obtain
\begin{equation}\label{dopplerani}
\frac{\ell^{2}C_{\ell}^{DD}}{2\pi} = \frac{T_{cmb}^{2}}{2\pi\ell^{2}}\int_{0}^{\eta_{0}}\deriv\eta\left(\frac{\partial}{\partial
\eta}\left[\dot{D}(-\dot\tau)
e^{-\tau}\right]\right)^{2}P\left(\frac{\ell}{r(\eta)}\right)(\eta_{0}-\eta)^{2}.
\end{equation}
Note that the amplitude of this expression is primarily determined by the factor
$\partial/\partial\eta(\dot D\dot\tau e^{-\tau})$, which as we saw
above depends on $\deriv/\deriv z\left[\bar{x}_{i}(z) H(z)\right]$;
the Doppler signal will be largest for models of reionization which
occur at high redshifts and which proceed quickly.

\section{The Reionization History}

%%%%%%%%%%%%% FIGURE 2: reionization histories and the homogenous part of the signal
\begin{figure*}[!t]
\begin{center}
\resizebox{8cm}{!}{\includegraphics{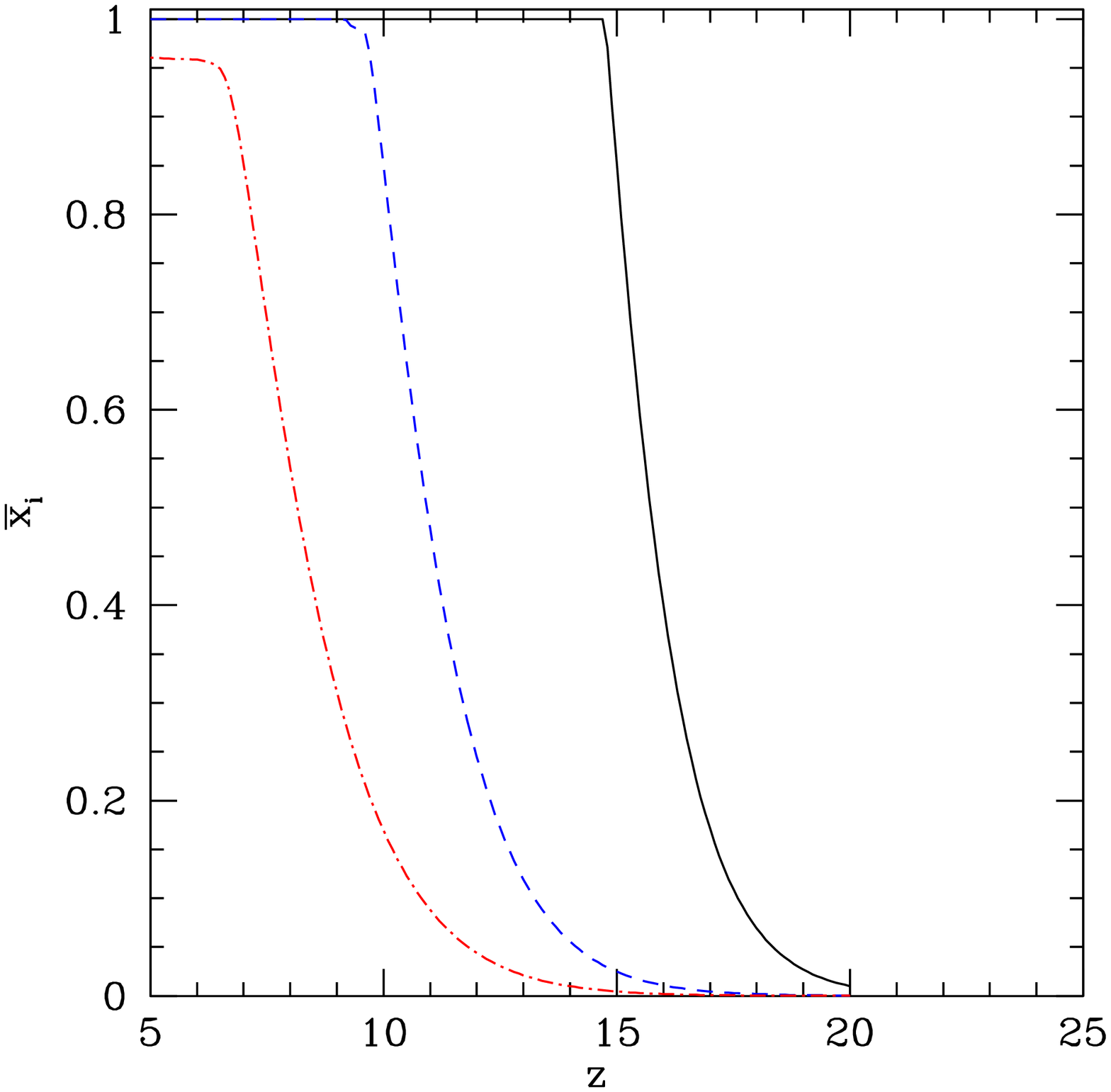}} \hspace{0.13cm}
\resizebox{8cm}{!}{\includegraphics{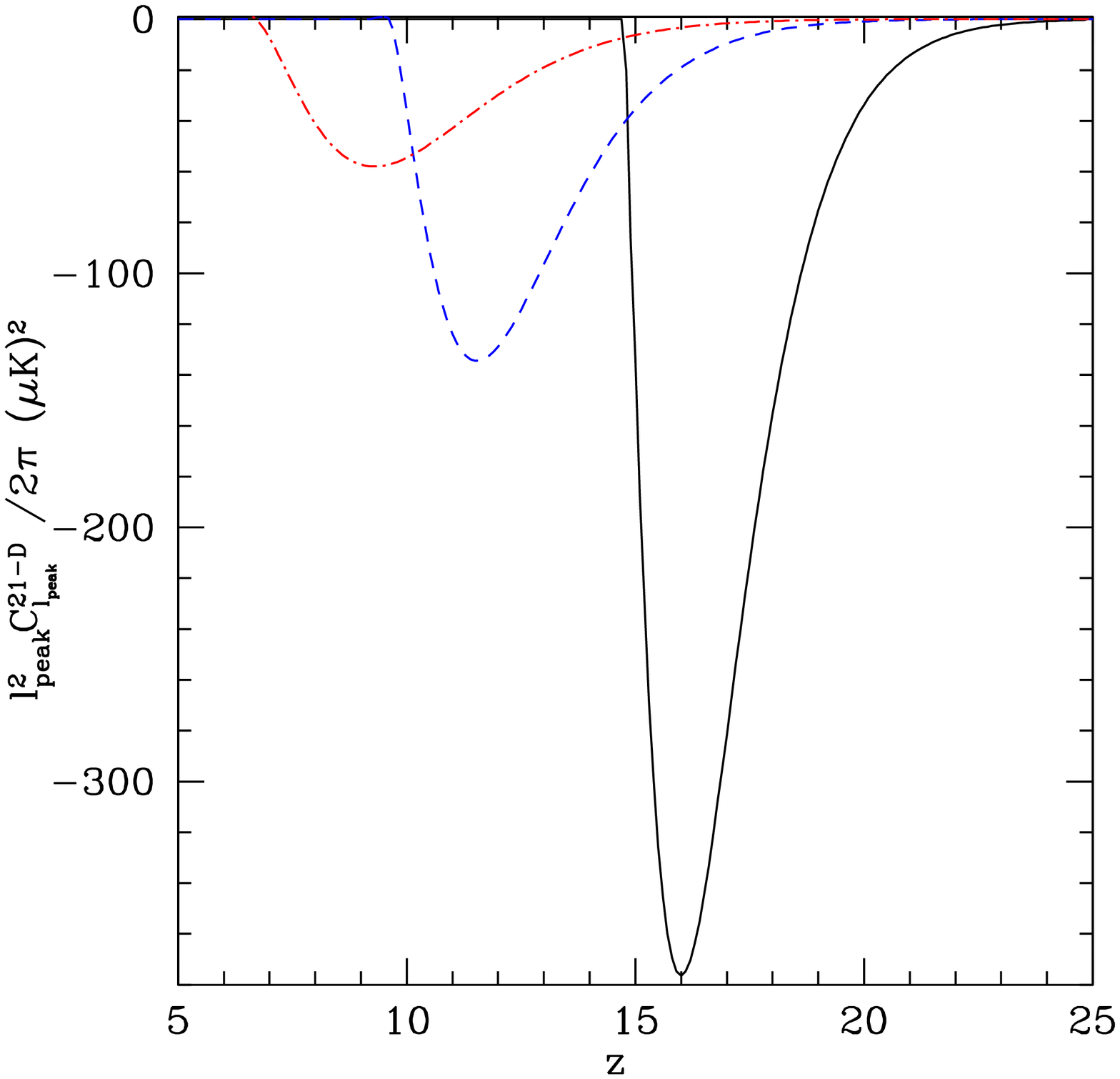}}
\end{center}
\caption{\emph{Left panel:} The reionization model of
\citet{Furlanetto:2006tf} with $\bar{x}_{i}\approx1$ at $z$=7
(dot-dashed curve), 10 (dashed curve ) and 15 (solid curve). \emph{Right
panel:} The contribution of the homogeneous term to the signal at
$\ell_{peak}$ for the same models.} \label{Profiles2}
\end{figure*}

The prescription of equation (\ref{model}) is not a well-motivated
model of reionization. The quantity of interest for the
cross-correlation, the rate at which the ionized fraction is
changing, is completely determined by the free parameter $\Delta z$, so the form has no predictive power.
Moreover, as soon as
the ionized fraction reaches 1/2, the rate of reionization begins to
slow, so the \emph{shape} of the signal is prescribed ``by hand" rather
than following from physically-motivated arguments (e.g., recombinations).  Our next task
is therefore to estimate the signal on more physically-satisfying
grounds.

However, in this paper we do not aim to present a complete model of
reionization. Instead we will use a simple model that encapsulates
many of the basic features in the global evolution. We begin by
assuming that a galaxy of mass $m_{\rm gal}$ can ionize a mass
$\zeta m_{\rm gal}$, where $\zeta$ is the ionizing efficiency: a
measure of the number of ionizing photons produced per baryon inside
galaxies. This amounts to the assumption of an ``inside-out"
reionization scenario wherein the gas is first ionized by galaxies
within overdense regions. This is consistent with numerical
simulations, at least when the sources resemble star forming
galaxies \citep{McQuinn:2005ce}. We then associate the reionization
rate with the star formation rate \citep{Furlanetto:2006tf};
\begin{equation}\label{ion}
\frac{\deriv\bar{x}_{i}}{\deriv t} = \zeta \frac{\deriv f_{\rm
coll}}{\deriv t} -
\alpha_{A}C(z,\bar{x}_{i})\bar{x}_{i}(z)\bar{n}_{e}(z),
\end{equation}
where $f_{\rm coll}$ is the mass bound to halos above some mass
threshold $m_{\rm min}$. In this work we take the minimum mass to
correspond to a virial temperature of $10^4$ K
\citep{Barkana:2000fd} and use the \citet{Press:1973iz} mass
function for simplicity. The second term on the right acts as a sink
describing recombinations. $\alpha_{A} = 4.2 \times 10^{-12}$
cm$^3$s$^{-1}$ is the case-A recombination coefficient at $10^4$ K,
$C\equiv \langle n^2_e\rangle/\langle n_e\rangle^2$ is the clumping
factor for ionized gas, and $\bar{n}_{e}$ is the average electron density in ionized
regions.

The physics underlying the ionization efficiency is highly uncertain
and depends on poorly understood intra-galactic dynamics, so in this
work we take it as a free parameter. In principle $\zeta$ and
$m_{\rm min}$ could vary with redshift due to feedback mechanisms.
In most cases, feedback decreases $\zeta$, and hence the
reionization rate. Thus in models that include feedback, we expect
the cross-correlation to be smaller and the peak to occur earlier in
the ionization history (e.g. \citealt{Furlanetto:2004nt}).
For example, if the photoheating that accompanies reionization
suppressed star formation in small galaxies, $\zeta$ would decrease
significantly once $\bar{x}_i \ga 0.5$, stretching reionization over
a longer time interval (possibly enough to create an inflection
point in the history; see below). In extreme cases, where
recombinations dominate early in reionization, there could even be a
turnover in $\bar{x}_i$ \citep{cen03,wyithe03-letter}, which would
strongly affect the cross-correlation \citep{Alvarez:2005sa}.

The clumping factor, $C$, can in principle be computed with
numerical simulations that incorporate radiative transfer. However,
this requires fully resolving small scale structure in the IGM as
well as tracking its evolution through reionization. Furthermore,
$C$ depends on how the IGM was ionized; whether low density gas was
ionized first or if many photons were consumed ionizing dense blobs
when $\bar{x}_{i}$ was small \citep{Furlanetto:2005xx}. This means
that $C$ must be recomputed for each different set of model
parameters. Simulations have unfortunately not yet reached these
goals. We instead  use a simple analytic model
\citep{Miralda-Escude:1998qs} in which voids are ionized
first.\footnote{This may at first seem to conflict with our
assumption that ionization begins in dense environments.  In fact,
the ionizing photons will begin in large-scale overdense regions,
but within the resulting ionized bubbles they will tend to ionize
the low-density gas first.  As a result, the
\citet{Miralda-Escude:1998qs} model underpredicts the recombination
rate by a factor of a few early in reionization
\citep{Furlanetto:2005xx}, but that has little effect on our
results.}  This model perhaps underestimates the mean recombination rate, and
so reionization probably proceeds a little slower that the model
predicts. This means that our models will overestimate the signal
and will predict a peak that is slightly later.
This optimistic assumption about the speed of reionization means
that our conclusions about the difficulty in observability of the
signal should be robust to uncertainties in the history of
reionization.

The left panel of Figure~\ref{Profiles2} shows the ionized fraction as
a function of redshift for three scenarios, with reionization
finishing by $z=15$, $z=10$ and $z=7$. They have CMB optical depths
of $\tau = 0.17$, $0.10$, and $0.06$ respectively.

We are now in a position to calculate the cross correlation under
the assumption that reionization is uniform ($\delta_{i}=0$
everywhere). In this case $P_{\delta i}(k) = 0$ and only the
first term in equation (\ref{21-Dopp}) contributes. The right panel
of Figure~\ref{Profiles2} shows $\ell^2C_{\ell}^{21-D}/2\pi$ for the
reionization models described above.
We show its value as a function of redshift at $\ell_{\rm peak}$;
the shape of the angular spectrum is only weakly-dependent on
redshift. Each curve corresponds to the matching reionization
history on the left. The signal is negative (implying an
anti-correlation) and peaks where the ionized fraction is changing
the fastest. The signal decreases toward lower redshifts for three
reasons. First and most importantly, the Hubble expansion dilutes
the electron densities and hence decreases the scattering rate of
CMB photons. Second, $\dot{f}_{\rm coll}/f_{\rm coll}$ decreases
with cosmic time (though only gently), so the relative ionization
rate decreases somewhat in scenarios with late reionization.
Finally, the clumping factor $C$ also increases with cosmic time,
slowing the tail end of reionization most severely when $z_r$ is
small.

\citet{Alvarez:2005sa} also used a physically-motivated model for
the ionization history that included both galaxy collapse and
recombinations. The major difference with our model is that they
prescribed $\bar{x}_i(z)$ to have a particular functional form that
matches smoothly onto $\bar{x}_i=1$ (something that we achieve with
a dynamically evolving clumping factor).  Comparing their Fig. 3
with the $z=15$ curve in our Figure~\ref{Profiles2}, we predict a
larger signal that peaks later in the reionization history. This
follows directly from the different reionization histories:  the
functional form prescribed by \citet{Alvarez:2005sa} has an
inflection point where $\bar{x}_{i}=1/2$, which causes their estimate to
peak slightly earlier. As noted above, such a feature could arise in
our model from feedback.  Our amplitude is also larger because
$\dot{f}_{\rm coll}/\fcoll$ is large throughout that epoch; again, the
smooth functional form prescribed by \citet{Alvarez:2005sa}
suppressed this dependence but may be more realistic in the presence
of strong feedback.

\subsection{Inhomogeneous Reionization}

Reionization is expected to be inhomogeneous for two reasons: the
clumpiness of the IGM and the clustering of the discrete ionizing sources.
The density-ionization cross correlation $P_{\delta i}$ is a measure of this
inhomogeneity. Given that little is known about the epoch of
reionization and that many of the parameters describing the
evolution are completely unconstrained, it is difficult to construct
detailed models for this correlation. We will therefore rely on the
simplest approach that we can.

To calculate this correlation we need a model of how the
perturbations to the ionized contrast $\delta_{i}$ grow through
reionization. Ideally we want to relate these to the underlying
matter overdensity $\delta$ so that we can express $P_{\delta i}$ in
terms of $P_{\delta\delta}$. One potential problem is that, on small scales, fully ionized bubbles
can have $\delta_i \gg 1$. Fortunately, on large scales these
bubbles average out so that linear theory will suffice in the
regime of interest.

As before, we assume reionization proceeds ``inside-out" so that a
galaxy of mass $m_{\rm gal}$ can ionize a bubble of mass $\zeta
m_{\rm gal}$ around it. It follows that, for an isolated large region
of overdensity $\delta$ \citep{Furlanetto:2004nh},
\begin{equation}\label{ionize}
x_{i}(\delta) = \bar{x}_{i}(1+\delta_{i}) = \zeta f_{\rm
coll}(\delta),
\end{equation}
where $f_{\rm coll}(\delta)$ is the conditional collapsed fraction
in our region. Note that the assumption of an \emph{isolated} region
is key here. It is adequate on the large, $\ga 100$ Mpc scales we
are interested in because ionizing photons are never able to travel
such large distances during reionization \citep{Furlanetto:2004nh}.
A model like that above fails if applied on $\sim 10 \Mpc$ scales,
because photons from sources outside the region become relevant.
Expanding this to linear order, we have $f_{\rm coll}(\delta) =
f_{\rm coll}(1+\bar{b} \delta)$. Here $\bar{b}$ is the mean galaxy
bias (averaged over all galaxies; \citealt{Mo:1995cs}). For an
individual galaxy of mass $m$, the \citet{Press:1973iz} mass
function yields
\begin{equation}
b(m) =
\left[1+\frac{\delta_{c}}{\sigma(m)^2D_{1}(z)}-\frac{1}{\delta_{c}(z)D_{1}(z)}\right],
\end{equation}
where $\delta_c$ is the critical density for spherical collapse and
$\sigma^2(m)$ is the fractional variance of the density field when
smoothed on spheres of mass $m$.  The average over all galaxies is
\begin{equation}
\bar{b} =\frac{1}{\fcoll} \int \deriv m~\frac{m}{\bar{\rho}} n(m)b(m).
\end{equation}
Note that the same expression follows from the reionization model of
\citet{Furlanetto:2004nh} by integrating over the galaxy population
of each ionized bubble in a large-scale region of overdensity
$\delta$. We show the mean bias as a function of redshift in the
upper left panel of Figure~\ref{fig: bias, fsat, crosscor}.

In this model, we therefore have $\delta_{i} = \bar{b} \delta$.
Before proceeding further, we must address two more potential
problems with our approach.  The condition that the ionized fraction
everywhere be less than unity implies
\begin{eqnarray}
1\geq x_{i} & = & \bar{x}_{i}(1+\bar{b} \delta),
\end{eqnarray}
which can be written
\begin{equation}
\bar{b} \delta\leq\frac{1-\bar{x}_{i}}{\bar{x}_{i}}.
\end{equation}
On the scales of interest we can estimate $\delta\sim\sigma(R)$.
$\ell=100$ corresponds to $R\sim 200$ Mpc, where $\sigma(200\Mpc,
z=10)\sim 0.004$. We are interested in the epoch in which
$\bar{x}_{i}\sim0.5$; so $\bar{b}$ can be as large as $\sim 360$ and
our approximations will still remain valid.

This prescription poses one final problem: as reionization proceeds,
the ionized contrast grows monotonically with the bias, irrespective
of the ionized fraction.  This ignores two subtleties of reionization. First, even on these
large scales the ionized features will eventually grow large enough
that the regions are no longer independent (i.e.,
low-density voids will eventually be ionized by their neighbors).
Moreover, \citet{Furlanetto:2005xx} have shown that recombinations
become important for larger bubbles. They imprint a maximum
size $R_{\rm max}$ on reionized regions; any extra photons produced
in these bubbles are canceled out by recombinations.
Although this process is difficult to model in detail (especially
given the weak constraints on structure in the IGM at high
redshift), we can approximately account for these effects by
modifying $\delta_i$ to
\begin{equation}\label{deltax}
\delta_i = \bar{b} \delta (1 - f_{\rm sat}),
\end{equation}
where $f_{\rm sat}$ is the fraction of the volume in the
``saturated" bubbles (with $R>R_{\rm max}$). As $\bar{x}_i$
approaches unity, so must $f_{\rm sat}$, so the effective bias
$\bar{b}(1-f_{\rm sat})$ eventually goes to zero.  In other words,
we subtract off photons produced inside regions where recombinations
dominate, because they are consumed before helping to ionize new
material.  Unfortunately $R_{\rm max}$ depends sensitively on the
distribution of high-redshift Lyman-limit systems, whose origins are
uncertain. There is thus no particularly good model for $R_{\rm
max}$ or $f_{\rm sat}$. However, the curves in Fig. 10 of
\citet{Furlanetto:2005xx} are reasonably well-described by
\begin{equation}
f_{\rm sat}(\bar{x}_{i}) = \exp\left[\frac{|\bar{x}_{i}-1|}{\Delta
\bar{x}_{i}}\right],
\end{equation}
where $\Delta\bar{x}_{i}$ is a parameter which encodes the effects
of the recombining bubbles (which increases as $R_{\rm max}$
decreases, because the saturation limit is reached earlier for
smaller bubbles). The upper right panel of Fig. \ref{fig: bias,
fsat, crosscor} shows $f_{\rm sat}$ for $\Delta \bar{x}_{i} = 0.2$,
$0.1$, and $0.05$. These values of $\Delta \bar{x}_{i}$ are chosen
to roughly match the results of \citet{Furlanetto:2005xx} and
correspond to $z=6$, $z=9$, and $z=12$ respectively (with
recombinations setting in earlier at lower redshifts because the
clumpiness $C$ increases with time).

Finally, the density-ionization cross correlation power spectrum is
given by:
\begin{eqnarray}\label{dens-ion}
 P_{i\delta}(k,z) = \nonumber
\langle\delta_{i{\bf k}}(z)\delta_{{\bf k}'}(z)\rangle & = &
\bar{b}(z)(1-f_{\rm
sat})\langle\delta_{{\bf k}}(z)\delta_{{\bf k}'}(z)\rangle \\
& = & \bar{b}(z)(1-f_{\rm sat})P(k, z).
\end{eqnarray}
Current simulations do not have large enough box sizes to test
this approximation on the $\ga 100$ Mpc scales of interest. However, on the
largest scales that we can compare ($k\sim0.1$ Mpc$^{-1}$), our result is $\sim 10\%$ larger than
that obtained from ``semi-numerical" simulations of the reionization process, at least early in reionization \citep{Mesinger:2007pd}.  As reionization proceeds, the simulations show that the amplitude on this scale gets even larger; this is because the characteristic size of ionized bubbles reaches $k\sim0.1$ Mpc$^{-1}$ about midway through the process.  We do not expect such an amplification to occur at the scales relevant to the cross-correlation so long as the bubbles do not reach such large sizes.

The lower panels of Figure \ref{fig: bias, fsat, crosscor} show the
variation in redshift space of the cross-correlation signal, with
the left panel showing the individual contributions of the homogeneous and inhomogeneous
terms and the right panel showing the sum. For reference, $z= 7$, 10
and 15 correspond to $\nu=$177.5, 129.1 and 88.8 MHz
respectively.

The bottom left panel of Figure \ref{fig: bias, fsat, crosscor}
shows that the positively correlated inhomogeneous part of the
signal strongly dominates over the anti-correlated homogeneous part.
This is because the inhomogeneous component is so strongly biased
(upper left panel).  Its amplitude increases rapidly with redshift
for two reasons:  the quantity $\deriv/\deriv z[H(z)\bar{x}_{i}(z)]$
is larger for earlier reionization and the galaxy bias grows large
at higher redshifts (see the upper left panel of Fig. \ref{fig:
bias, fsat, crosscor}). Our amplitudes are somewhat larger than
those of \citet{Alvarez:2005sa} because our reionization models
allow larger $\deriv \bar{x}_i/\deriv z$ and because our cosmology
yields larger bias factors at high redshift.

In the lower right panel we explore the effects of varying the rate
at which bubbles saturate by varying the parameter $\Delta
\bar{x}_{i}$. If saturation sets in earlier the peak amplitude of
the signal decreases, and the peak also moves earlier in
reionization.  However, in all of these cases $f_{\rm sat}$ has
little effect except in a brief redshift interval. This is simply
because $\bar{x}_i$ increases from $\sim 0.5$ to $\sim 1$ so
quickly. We find, therefore, that measuring the peak amplitude of
the cross correlation does not precisely yield the rate at which
reionization is occurring (or, in our simple model, roughly the rate
at which gas collapses onto galaxies); rather, it provides the rate
at which photons are able to escape from the environs of their
source galaxy.  Late in reionization, the overall amplitude will be
suppressed by the recombining bubbles that host most of the ionizing
sources.  Recovering $\bar{x}_i(z)$ will require carefully modeling
this effect.

When inhomogeneous reionization is included, our models robustly
predict a positive correlation between the CMB and 21 cm signals
(see also \citealt{Alvarez:2005sa}).  This follows from the strong
clustering of the first galaxies and reflects our assumption of
``inside-out" reionization.  An anti-correlated signal would imply
that low-density gas is ionized at least as quickly as high-density
gas; thus the overall sign of the correlation would provide a clean
test of these contrasting approaches to reionization.

%%%%%%%%%%%%% FIGURE 3: Galaxy bias, saturated fraction, split inhomogeneous and homogeneous and total
\begin{figure*}[!t]
\begin{center}
\resizebox{8cm}{!}{\includegraphics{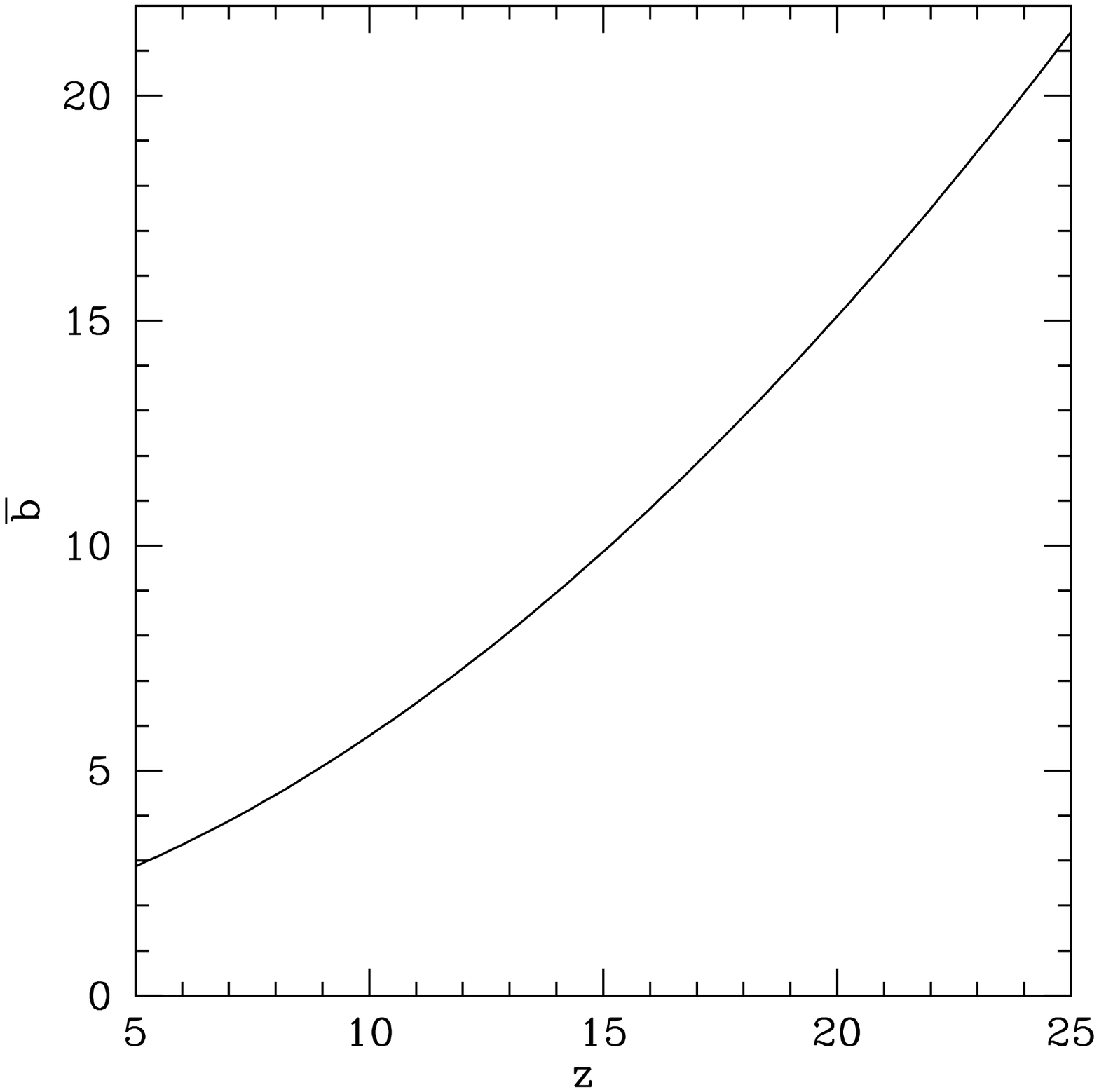}} \hspace{0.13cm}
\resizebox{8cm}{!}{\includegraphics{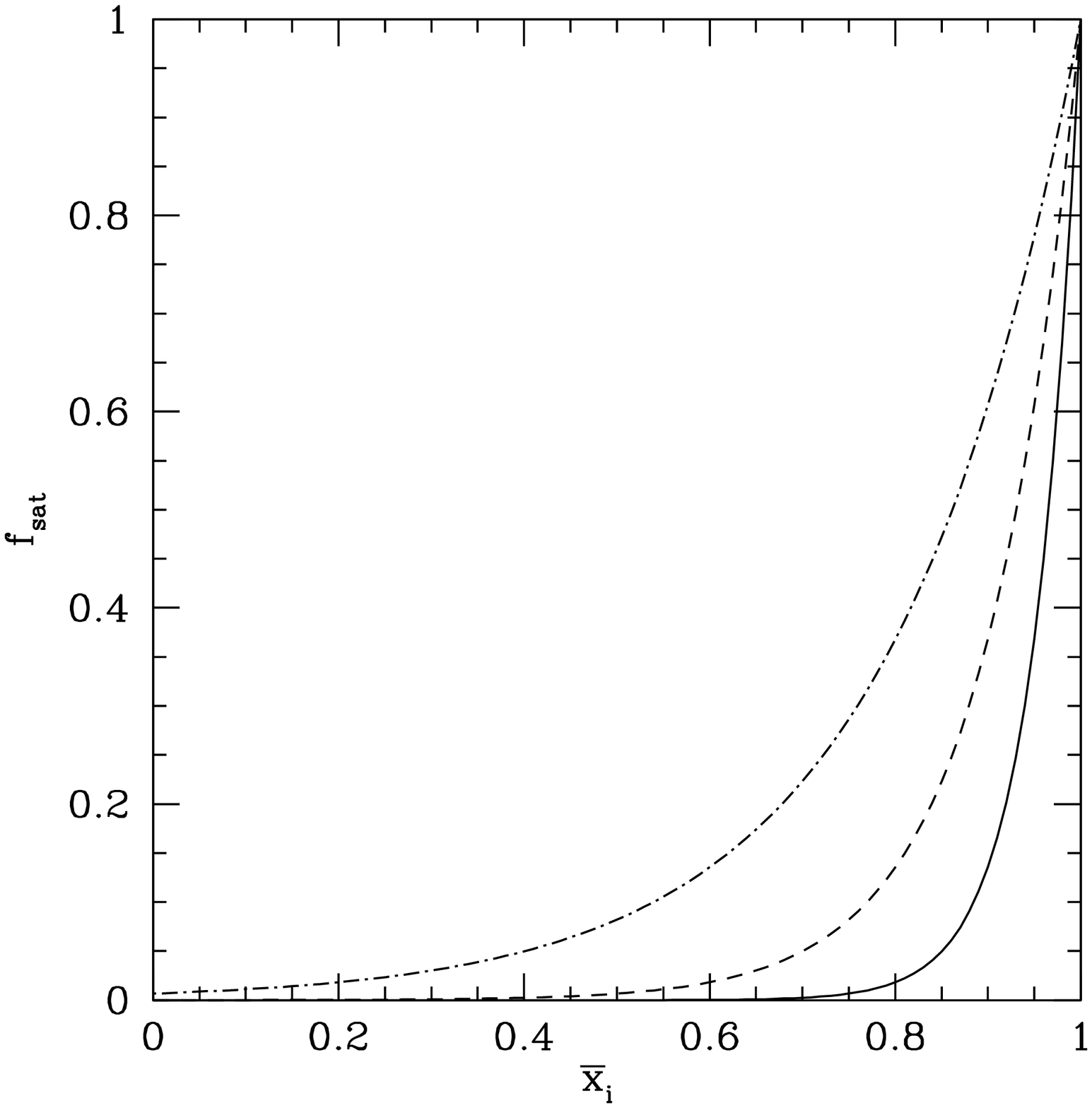}}\\
\resizebox{8cm}{!}{\includegraphics{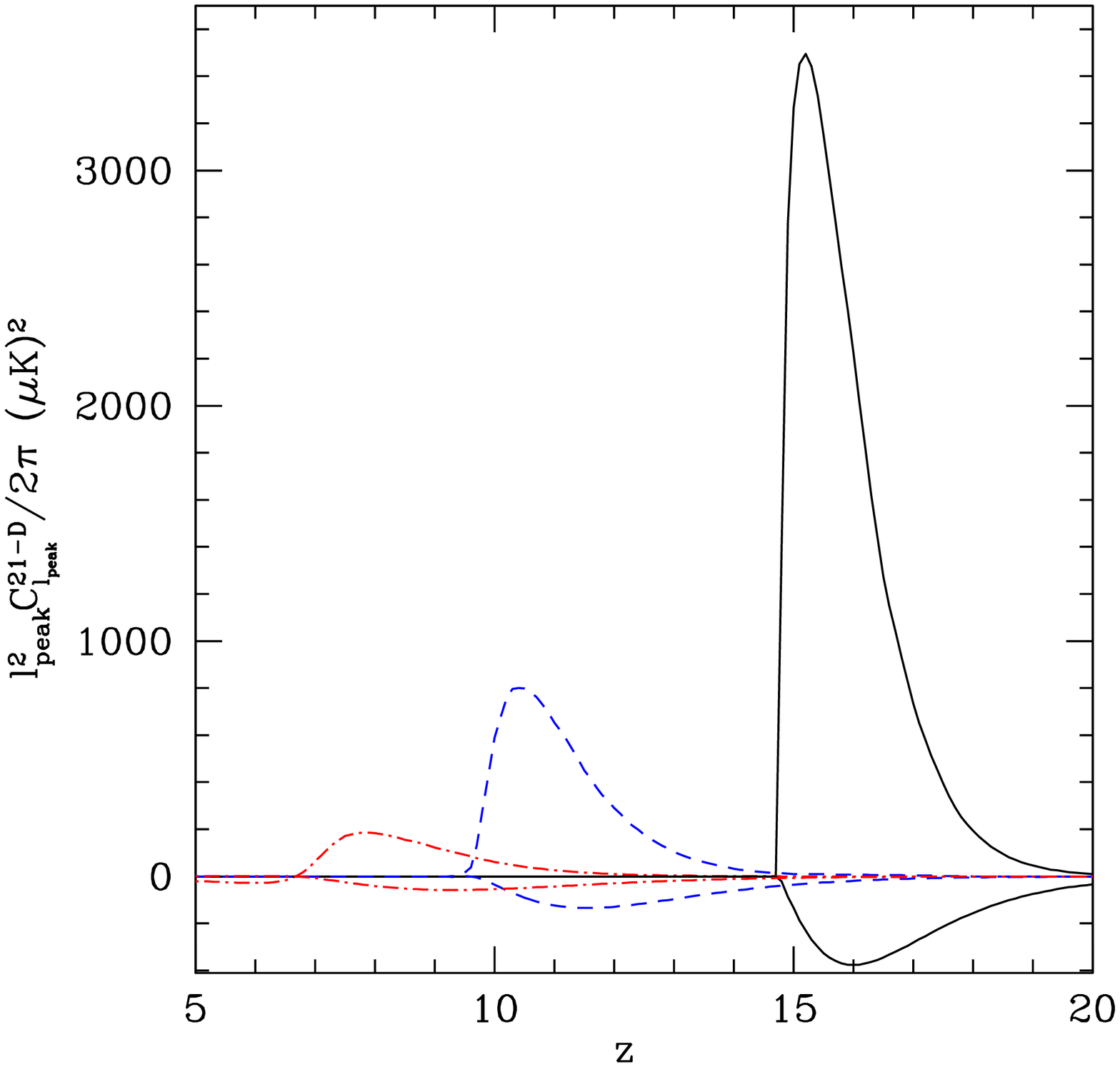}} \hspace{0.13cm}
\resizebox{8cm}{!}{\includegraphics{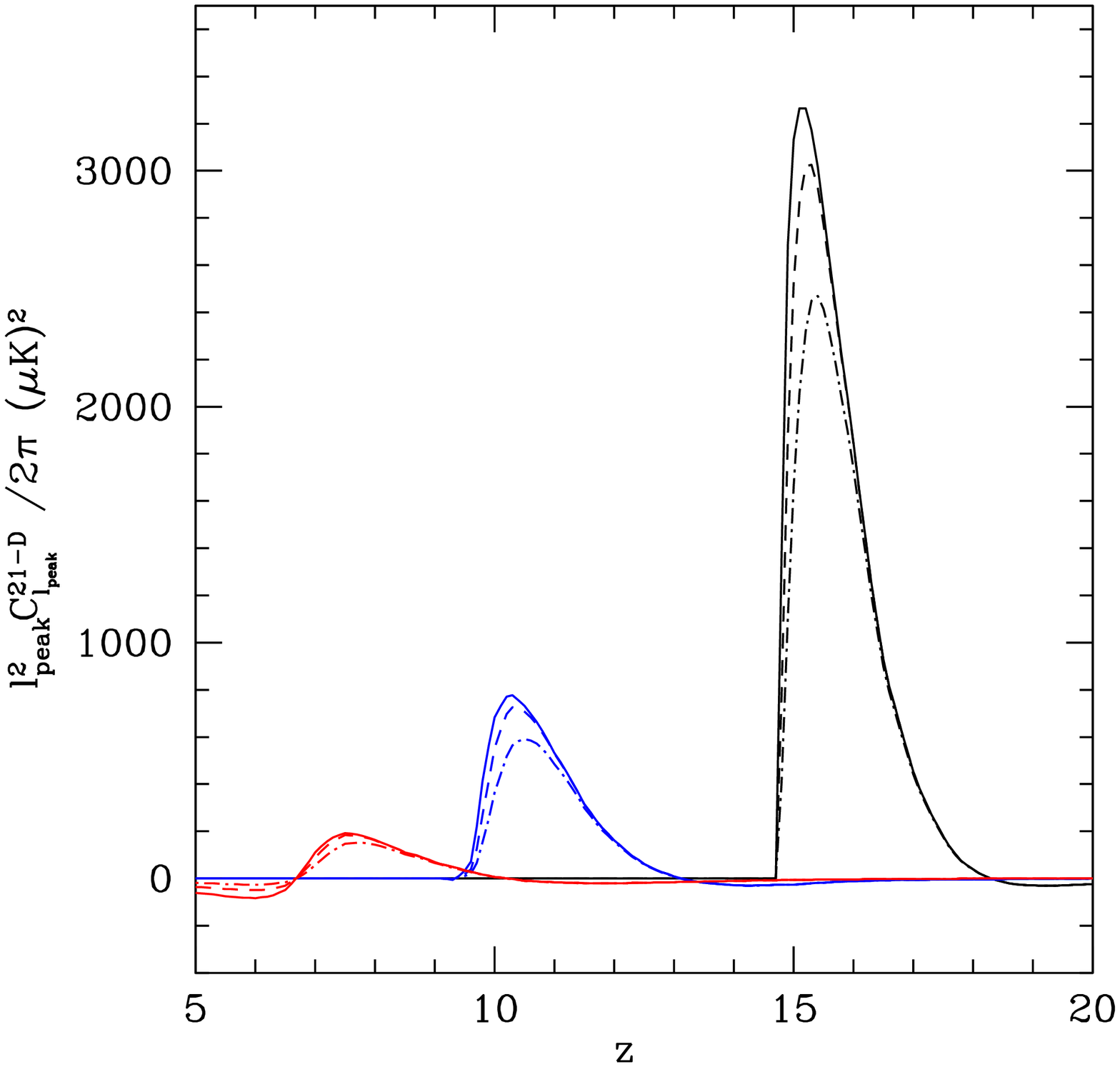}}
\end{center}
\caption{\textit{Upper Left:} The mean galaxy bias function.
\textit{Upper Right:} $f_{\rm sat}$ for $\Delta\bar{x}_{i} = 0.05$
(solid curve),  $0.1$ (dashed curve), and $0.2$ (dot-dashed curve).
\textit{Bottom left:} The homogeneous and inhomogeneous
contributions to the cross correlation at $\ell_{peak}$
corresponding to reionization models with $\bar{x}_{i}$=1 at
$z\approx7$, 10 and 15. In each case, the positive (negative) curves
show the contribution from $P_{\delta i}$ ($P$).  All curves assume
an $f_{\rm sat}$ with $\Delta \bar{x}_{i}=0.05$. The corresponding
reionization histories are shown in Fig. \ref{Profiles2}.
\textit{Bottom right:} The total signal at $\ell_{peak}$. At each
$z$, we show results for the different values of the parameter
$\Delta\bar{x}_{i}$ in $f_{\rm sat}$ shown above.} \label{fig: bias,
fsat, crosscor}
\end{figure*}

\section{Foregrounds and Observability}

There are two unavoidable sources of noise in any attempt to measure cosmic
signals:  the intrinsic noise associated with the detector and
the error associated with sample variance (especially at low multipoles). The
latter results from having only one sky and only $2\ell+1$
independent samples at each $\ell$:
\begin{equation}
\Delta C_{\ell}^{sv} = \sqrt{\frac{2}{(2\ell+1)f_{\rm
sky}}}C_{\ell},
\end{equation}
where $f_{\rm sky}$ is the fraction of sky being observed. For a
single 21 cm observing field this is given by $f_{sky}=d\Omega/4\pi
\approx \lambda^2/(A_{d}4\pi)$, approximated by the diffraction
limit of the telescope, where $A_{d}$ is the area of a single dish
and $\lambda = 21(1+z)$ cm.

\begin{figure}
\begin{centering}
  % Requires \usepackage{graphicx}
  \includegraphics[width = 8 cm]{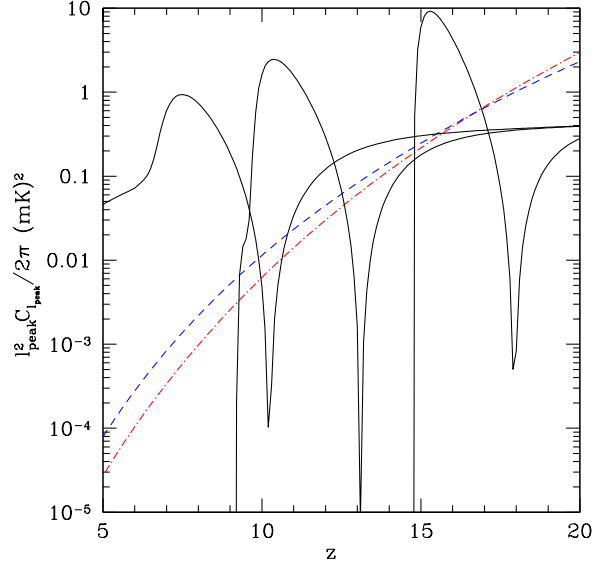}\\
  \caption{The thermal detector noise for 21 cm observations with the MWA (dot-dashed curve) and the SKA (dashed curve) and the
  intrinsic 21 cm signal for the reionization models in Fig. \ref{Profiles2} (solid curves).} \label{errors}
  \end{centering}
\end{figure}

While the cosmic variance contribution is purely geometric, the
detector noise depends on the details of the observation.  For 21 cm
experiments, the RMS detector noise fluctuation per visibility of an
antennae pair observing for a time $t_{0}$ in one frequency channel
is \citep{McQuinn:2005hk}
\begin{equation}
\Delta V^{N} = \frac{\lambda^{2}T_{\rm sys}}{A_{d}\sqrt{\Delta \nu
t_{0}}},
\end{equation}
where $T_{\rm sys}(\nu)$ is the total system temperature.
An irreducible limit on $T_{\rm sys}$ is provided by the radio sky itself, so
$T_{\rm sys}\geq T_{\rm sky}$. In fact, at the low frequencies
relevant to the redshifted 21 cm sky, the noise is strongly
dominated by synchrotron electrons from fast electrons in the
Milky Way, so to a good approximation $T_{\rm sys}\approx T_{\rm
sky}$. A good rule of thumb for the sky temperature of high
latitude, ``quiet" portions of the sky is \citep{Furlanetto:2006jb}
\begin{equation}
T_{\rm sky} \sim 180\left(\frac{\nu}{180\textrm{MHz}}\right)^{-2.6}
\textrm{K}.
\end{equation}
It follows that the detector noise in the auto-correlation for a
single baseline is
\begin{equation}
C_{1b}^{N} = \left(\frac{\lambda^{2}T_{\rm
sys}}{A_{d}}\right)^{2}\frac{1}{\Delta \nu t_{0}}.
\end{equation}
The time spent observing each visibility is
\begin{equation}
t_{{\bf u}} = \frac{A_{d}t_{0}}{\lambda^{2}}n({\bf u}),
\end{equation}
where $u = \ell/2\pi$ and $n(\bf u)$ is the average number of
baselines that can observe the mode ${\bf u}$ at any instant,
normalized so that its integral over the ${\bf u}$ plane equals the
total number of baselines in the interferometer
\citep{McQuinn:2005hk}. The covariance matrix for an interferometer
is then
\begin{equation}
C^{21,N}({\bf u}) = \left(\frac{\lambda^{2}T_{\rm
sys}}{A_{d}}\right)^{2}\frac{1}{\Delta \nu t_{\bf u}}.
\end{equation}
The total error on the autocorrelation is a combination of
sample variance and detector noise
\begin{equation}
\Delta C_{\ell}^{21} = \sqrt{\frac{2}{(2\ell+1)f_{\rm sky}}}\left\{
\left[ \left(\frac{\lambda^{2}}{A_{d}}\right)^{3}\frac{T_{\rm
sys}^2}{\Delta \nu t_{0}n(\bf
u)}\right]^{2}+\left(C_{\ell}^{21}\right)^{2}\right\}^{1/2}.
\end{equation}
We estimate $C_\ell^{21}$ using equation~(\ref{cl21}) and the
inhomogeneous reionization model described above; note that
$P_{ii}(k)=\bar{b}^2 (1-f_{\rm sat})^2 P(k)$ on these large scales.
Figure~\ref{errors} shows the intrinsic 21 cm signal for our fiducial
reionization models as well as the thermal detector noise,
$\ell^{2}C_{\ell}^{21,N}$ for two planned experiments,
the Mileura Widefield Array-Low Frequency Demonstrator (MWA) and the
Square Kilometer Array  (SKA).  We have assumed a frequency window
of $\Delta\nu = 1\MHz$, 1000 hours of observing time, and $\ell =
100$. The area of a single dish is taken to be the minimum baseline:
$(4~\rm m)^{2}$ for MWA and $(16~\rm m)^{2}$ for the SKA, and we
assume a filled core outside of which the density of antennae falls like
$r^{-2}$.  The total effective areas are $1$ km$^2$ for the SKA and
$\sim 7 \times 10^{-3}$ km$^2$ for the MWA.

The 21 cm signal falls sharply early in reionization because of the
prefactor on the power spectrum, $[\bxhi(z)-\bar{b}(1-f_{\rm
sat})\bar{x}_{i}(z)]^{2}$. These two terms cancel at a point early
in reionization when the increased total density matches the
decreased ionized fraction; at later times, the inhomogeneous term
dominates and the signal increases again.  Obviously, except when
reionization occurs at $z=15$ the intrinsic 21 cm signal
significantly dominates the thermal noise. Thus any measurement of
the cross-correlation is sample variance limited. Note the
difference with typical 21 cm experiments, which are usually
dominated by thermal noise:  in this case, we are interested in such
large scales that the 21 cm telescopes can measure the modes in
detail at least neglecting foreground contamination). We neglect the
21 cm detector noise for the remainder of the paper.

%Figure 4: Thermal Detector noise and intrinsic 21 cm power.

For the CMB, experiments like \emph{WMAP} are already nearly
signal-variance limited on scales near $\ell_{peak}$.  Thus we
neglect the thermal noise term entirely for the CMB.

One of the general advantages of cross correlation is that any
uncorrelated noise (such as the thermal noise in each detector) will
vanish.  Unfortunately, many of the strong foregrounds for 21 cm
observations also appear at CMB frequencies (such as the Galactic
synchrotron radiation and radio point sources).  However, we will
assume optimistically that cleaning of the separate maps will
effectively remove this contamination \citep{Zaldarriaga:2003du,
morales04,santos03,morales05_foregrounds,McQuinn:2005hk}; in reality,
foregrounds could make the signal even more difficult to detect than
estimated here.

We can then estimate the 1-$\sigma$ error for the cross-power spectrum from
\begin{equation}\label{noise}
(\Delta C_{\ell}^{21-D})^{2}\approx \frac{1}{(2\ell+1)f_{\rm
sky}\Delta\ell}\left[(C_{\ell}^{21-D})^{2}+ C_{\ell}^{CMB}
C_{\ell}^{21}\right],  %+C_{\ell}^{21,~N})\right],
\end{equation}
where $\Delta\ell$ is the size of the bins over which the power spectrum
is averaged ($\ell-\Delta\ell/2<\ell<\ell+\Delta\ell/2$). In
terms of the cross-correlation coefficient, the expected fractional
error (eq.~\ref{noise}) can be
%SF:  added footnote
written:\footnote{Note that our definition of the cross-correlation coefficient $r$ ignores the noise terms.}
\begin{equation}
\frac{\Delta C_{\ell}^{21-D}}{C_{\ell}^{21-D}} =
\frac{1}{\sqrt{(2\ell+1) f_{\rm sky}\Delta
\ell}}\frac{\sqrt{1+r^{2}}}{r}.
\end{equation}
This result has a simple physical interpretation. The first factor
on the right hand side is the fractional error from cosmic
variance for any autocorrelation.  If the two signals are weakly
correlated, $r \ll 1$, only a portion of each intrinsic signal is
useful for the measurement and the cosmic variance limit increases
accordingly.  This can be recast as a constraint on the minimum
correlation coefficient $r$ observable with a given signal-to-noise
on the cross-power:
\begin{equation}\label{correlation}
r > \left[ \left(\frac{\Delta
C_{\ell}^{21-D}}{C_{\ell}^{21-D}}\right)^{2}(2\ell+1)f_{\rm
sky}\Delta\ell- 1\right]^{-1/2}.
\end{equation}

Thus the observability of our signal will depend on the
cross-correlation coefficient. We can estimate its value by
comparing equation~(\ref{dopplerani}) for our reionization models to
the total primordial CMB anisotropies.  For the models
in Figure~\ref{Profiles2}, we find $\ell^{2}C^{DD}_{\ell}/2\pi =
19.8~\mu$K$^{2}$, $2.4~\mu$K$^{2}$, and $0.4~\mu$K$^{2}$ for the
$z=15$, $z=10$, and $z=7$ cases respectively. The primary anisotropy
is $\approx (40~\mu\rm K)^2$ at $\ell = 100$, so even if the Doppler
component of the CMB is completely correlated with the $21$ cm
signal (see below), we must still have $r<0.11$, $0.04$ and $0.015$
for the $z=15$, $10$ and $7$ cases,  respectively. One can already
see that the prospects for detecting the cross-correlation are dim:
the CMB at $\ell \approx 100$ is well-fit by the primary
anisotropies alone, ignoring the Dopper terms from reionization, and
has been observed to nearly the cosmic variance limit
\citep{Spergel:2006hy}.  The fractional cosmic variance errors for
$C_\ell^{21-D}$ will be larger than those on the CMB by a factor
$1/r$.

For a concrete (though optimistic) estimate, let us assume that we
have $f_{\rm sky}=1/2$ and sum over all multipoles. Thus we forsake
any shape information in order to test the null hypothesis, that is,
whether or not correlation can be distinguished from no correlation.
For this calculation, we assume for simplicity that the correlation
coefficient between the Doppler anisotropy and the 21 cm brightness,
$r_{21-D} = r \sqrt{C_\ell^{CMB}/C_\ell^{21-D}}$, is independent of
scale and we choose the observed frequency to correspond to the peak
of the cross-correlation. We can then find the total signal to noise
ratio from (compare eq.~51 of \citealt{Slosar:2007sy})
\begin{equation}
SNR^{2} = \sum_{\ell}\left(\frac{C_{\ell}^{21-D}}{\Delta
C_{\ell}^{21-D}}\right)^{2} \approx r_{21-D}^{2}\sum_{\ell}(2\ell+1)f_{\rm
sky}\frac{C^{DD}_{\ell}}{C^{cmb}_{\ell}},
\end{equation}
since $C^{DD}_{\ell}/C^{cmb}_{\ell}$ is small. We find signal to
noise ratios of $SNR \sim 2\,r_{21-D}$, $5\,r_{21-D}$, and
$14\,r_{21-D}$ for reionization ending at $z=7,\,10,$ and $15$,
respectively. If the 21 cm signal and the Doppler signal were
perfectly correlated, then we could conceivably detect the
correlation in all these scenarios. However, in reality the
correlation coefficient between the 21 cm signal and the Doppler
signal is $r_{21-D} \sim 0.3$. Thus it is unlikely that we can
detect the cross-correlation if reionization ends at $z \la 12$. For
reionization at higher redshifts, a detection is possible provided
we can view $\ga 50\%$ of the sky.

Of course, the above estimate assumes a single frequency band for
the 21 cm map; increasing the frequency coverage will provide more
samples and improve the total signal-to-noise (though at the cost of
losing any information about the signal's evolution with redshift).
However, $\ell \sim 100$ corresponds to $200 \Mpc$.  Demanding that
each 21 cm frequency slice be statistically independent then
requires that the slices be $\sim 10 \MHz$ thick, which corresponds
to $\Delta z \sim 1$.  Thus there are in fact relatively few
independent slices available, unless reionization is considerably
more extended than in our simple models (in which case each
individual slice would have a smaller signal anyway).

\section{Discussion and Conclusion}

In this work we have calculated the large scale cross correlation
between the CMB temperature anisotropy and the 21 cm
background, improving the calculation of \citet{Alvarez:2005sa}. The
signal is expected to come from large angular scales ($\ell\sim100$)
corresponding to $\sim 200$ Mpc at redshifts $z\sim10$. On these
scales linear theory is still valid, which greatly simplifies the
analysis. We have also presented a new model for the evolution of
the ionized contrast as reionization proceeds that includes the
effects of recombinations in limiting the apparent bias of reionized
regions.

The cross correlation arises from the connection between linear
overdensities and the baryon velocity field via the continuity
equation in linear theory. Fluctuations in the underlying density
field source brightness fluctuations in the 21 cm background, while
the baryon velocity field Doppler scatters the CMB. The evolving
ionized fraction, which is analogous to the evolving gravitational
potentials in the integrated Sachs-Wolfe effect, reduces the line of
sight cancelation that is usually associated with Doppler
contributions to secondary anisotropies. Moreover, the redshift
information contained in the 21 cm signal (as a binning in
frequency) allows us to reconstruct 3D information about the cross
correlation.

Our calculations include improvements to the approach originally
presented by \citet{Alvarez:2005sa}. The key differences are the
treatment of the bulk velocity corrections to the 21 cm signal, the
reionization model, and the ionized fraction-density
cross-correlation.

\citet{Alvarez:2005sa} chose a particular functional form $\ln
[1-\bar{x}_{i}] = -\zeta_{0}(z)f_{\rm coll}(z)$ for the reionization
history.  Although it is driven by the same mechanism as ours early
in reionization (the time evolution of $\fcoll$), this form demands
that the reionization history slow when $\bar{x}_i > 0.5$.  Although
such a delay can be loosely attributed to either feedback or
recombinations, there is in fact no requirement that they set in so
early, or that they slow the subsequent evolution so dramatically.

Our model, on the other hand, explicitly relates the collapsed
fraction to the ionized fraction, to which we add the effects of
recombinations (which limit the bubble sizes and mediate the source
bias). Following \citet{Furlanetto:2005xx}, we assume that
ionized bubbles grow until they reach a size $R_{\rm max}$, where
recombinations prevent subsequent growth. Any more ionizing photons
produced in these bubbles are canceled out by recombinations. As
reionization proceeds more and more bubbles saturate and no longer
contribute to the global evolution of the ionized fraction. As this
process unfolds, more of the collapsed matter enters these saturated
bubbles, so the ionized contrast between over- and underdense
regions begins to get smaller, eventually shrinking to zero as all
regions become ionized.

These differences change the amplitude and evolution of the signal
but not its qualitative behavior. Like \citet{Alvarez:2005sa}, in
the case of homogeneous reionization we find an anti-correlation
between the 21 cm signal and the CMB. In this case an overdensity
simultaneously increases the brightness of the 21 cm background and
cools the CMB through Thomson scattering across regions where
$x_{i}$ is increasing. The contribution from the homogeneous term
decreases with redshift due to dilution of the matter density
through Hubble expansion.

The result is markedly different when inhomogeneous reionization is
included: the 21 cm and CMB signals become strongly correlated.
Ionized clumps still scatter and cool the CMB photons through the
$x_{i}$ gradient. However, these regions also cause a significant
\textit{negative} fluctuation in the 21 cm brightness due to the
large deficit of neutral hydrogen. This term is strongly boosted by
the galaxy bias and dominates over the anti-correlated homogeneous
reionization contribution.

Of course, this positive correlation rests on our assumption that
reionization proceeds inside-out. If instead low density gas is
ionized first, the inhomogeneous component would become
anti-correlated: the underdense regions would blueshift the CMB
photons while decreasing the 21 cm brightness both through a
decreased matter density and ionized fraction.

Unfortunately, we have shown above that the 21 cm and CMB signals
are too weakly correlated for the cross-correlation to be easily
detectable.  Like \citet{Alvarez:2005sa}, we find that thermal noise
will not pose a significant problem because the correlation only
appears on large scales, where 21 cm telescopes are sample
variance-limited.  However, when we include cosmic variance of the 21 cm signal, the
prospects for detection look dim.
The cross-correlation coefficient between the total CMB signal and
the 21 cm background is only $r \la 0.03$:  it is small because the
Doppler contribution makes up $\la 1\%$ of the total CMB anisotropy,
and only $\sim 30\%$ of that correlates with the 21 cm background.
Experiments targeting this signal will require extremely large
fields of view, and even then will be unable to recover the signal
at high signal-to-noise or to trace its angular power spectrum.
Prospects are actually best for early reionization, because then the
CMB Doppler term is increased by the denser gas during reionization.
(Note also that we have ignored foreground contamination in the 21
cm signal, which must be cleaned and will further degrade the
detectability.)

Although it is possible that our models under-predict the
cross-correlation signal, our predictions for the detectability is
robust. We have shown that the signal to noise ratio is limited
primarily by the smallness of the Doppler signal. The size of the
Doppler contributions is determined entirely by the rate of
reionization and the epoch at which reionization occurs. Our
models already provide relatively fast reionization, so the signal can probably only
be increased in this fashion by a factor of $\sim 2$, which
does not change our conclusions. Furthermore, while detection of the
cross-correlation at $\ell\sim100$ is limited by the cosmic variance
of the 21 cm and CMB signals, it does not help to move to higher
multipoles where there are more modes available.  On the one hand, moving to higher $\ell$ may increase the effective bias, $\bar{b}(z)$, and thus the cross-correlation, as the scale approaches the characteristic bubble size.  However, this aspect would only boost the 21 cm signal (which does not help much because it also increases the cosmic variance) and not the Doppler
signal.  In fact, as shown by \citet{Giannantonio:2007za} the Doppler signal, and hence the ratio
$C_{\ell}^{DD}/C_{\ell}^{CMB}$, begins to sharply decrease after
$\ell\sim100$: any gains made in sampling more modes are immediately
lost.

We thank M. Zaldarriaga and M. Alvarez for helpful comments on the
manuscript and A. Mesinger for sharing simulation data with us.

\bibliographystyle{mn2e}
\bibliography{pap}{}

\appendix
\section{Integral Approximation}
Consider the integral representation of the 3D Dirac delta function,
\begin{equation}
(2\pi)^{3}\delta({\bf r}-{\bf r}')=\int \deriv^{3}k \exp[i {\bf
k}\cdot({\bf r}-{\bf r}')].
\end{equation}
Expanding the plane wave in spherical coordinates, we obtain
\begin{eqnarray}
(2\pi)^{3}\delta({\bf r}-{\bf r}')& = & (4\pi)^{2} \int \deriv^{3}k
\sum_{\ell, m}(-i)^{\ell}j_{\ell}(kr)Y_{\ell m}(\hat{{\bf
k}})Y_{\ell m}^{*}(\hat{{\bf r}}) \sum_{\ell',
m'}(i)^{\ell'}j_{\ell'}(kr')Y_{\ell'
m'}^{*}(\hat{{\bf k}})Y_{\ell' m'}(\hat{{\bf r}'}) \\
& = & \label{a2}(4\pi)^{2}\sum_{\ell,m}\int k^{2}\deriv k
j_{\ell}(kr) j_{\ell}(kr')Y_{\ell m}(\hat{\bf r})Y_{\ell
m}^{*}(\hat{\bf r}'),
\end{eqnarray}
where we have used the orthogonality of the spherical harmonics. Now
in spherical coordinates we can express the Dirac delta function as
\begin{equation}
\delta({\bf r}-{\bf r}')= \delta(\Omega,
\Omega')\frac{\delta(r-r')}{r^{2}},
\end{equation}
so that multiplying equation (\ref{a2}) by $Y_{\ell m}(\hat{\bf r}')$
and integrating over $d\Omega '$ yields
\begin{equation} \label{app bess ident}
\int_{0}^{\infty}k^{2}\deriv k j_{\ell}(kr)j_{\ell}(kr')=
\frac{\pi}{2}\frac{\delta(r-r')}{r^{2}}.
\end{equation}
In this paper we encounter an integral similar to the left hand side
of equation (\ref{app bess ident}), but with another smooth function
of $k$, $P(k)/k^2$, in the integrand. The spherical Bessel functions
$j_{\ell}(kx)$ are very small for $kx<\ell$ and start to oscillate
for $kx \sim \ell$. As can be seen from (\ref{app bess ident}), the
Bessel functions are out of phase for any separation of the points
$r$ and $r'$. The integral will thus only receive contributions from
a region around the first peak, which occurs at $k \sim \ell/x$. We
then make the approximation $P(k)\approx P(k=\ell/r)$ (e.g.,
\citealt{Zaldarriaga:2003du}), and pull it out of the integral to
obtain the approximation used in the main body of the paper:
\begin{equation} \label{alverez approx}
    \frac{2}{\pi}\int_{0}^{\infty}k^2{}\deriv k \frac{P(k)}{k^{2}}j_{\ell}(kr)j_{\ell}(kr')\approx
    P\left(k=\frac{\ell}{r}\right)\frac{\delta(r-r')}{\ell^{2}}.
\end{equation}
\end{document}